%% file: main.tex
\documentclass[conference]{IEEEtran}
\IEEEoverridecommandlockouts
\usepackage{cite}
\usepackage{amsmath,amssymb,amsfonts}
\usepackage{graphicx}
\usepackage{textcomp}
\usepackage{xcolor}

\usepackage[export]{adjustbox}

\usepackage{amssymb}
\usepackage{amsmath}
\usepackage{pgfplots}

\usepackage[hyphens]{url}

\usepackage{multirow}
\usepackage[ruled,vlined]{algorithm2e}

\usepackage{subfigure}		
\usepackage{tikz}

\usepackage{xcolor}

\usepackage{xcolor,pifont}
\newcommand*\colourcheck[1]{%
  \expandafter\newcommand\csname #1check\endcsname{\textcolor{#1}{\ding{52}}}%
}
\colourcheck{blue}
\colourcheck{green}
\colourcheck{red}

\newcommand*\colourx[1]{%
  \expandafter\newcommand\csname #1x\endcsname{\textcolor{#1}{\ding{54}}}%
}
\colourx{red}

\def\BibTeX{{\rm B\kern-.05em{\sc i\kern-.025em b}\kern-.08em
    T\kern-.1667em\lower.7ex\hbox{E}\kern-.125emX}}

\begin{document}

\title{Efficacy of Satisfiability-Based Attacks in the Presence of Circuit Reverse-Engineering Errors\\
}

\author{
\IEEEauthorblockN{Qinhan Tan}
\IEEEauthorblockA{\textit{College of Computer Science and Technology} \\
\textit{Zhejiang University}\\
Hangzhou, China \\
tanqinhan@zju.edu.cn}
\and
\IEEEauthorblockN{Seetal Potluri}
\IEEEauthorblockA{\textit{Department of ECE} \\
\textit{North Carolina State University}\\
Raleigh, U.S. \\
spotlur2@ncsu.edu}
\and
\IEEEauthorblockN{Aydin Aysu}
\IEEEauthorblockA{\textit{Department of ECE} \\
\textit{North Carolina State University}\\
Raleigh, U.S. \\
aaysu@ncsu.edu}
}

\maketitle

\begin{abstract}
%

Intellectual Property (IP) theft is a serious concern for the integrated circuit (IC) industry. To address this concern, logic locking countermeasure transforms a logic circuit to a different one to obfuscate its inner details. 
The transformation caused by obfuscation is reversed only upon application of the programmed secret key, thus preserving the circuit's original function.
This technique is known to be vulnerable to Satisfiability (SAT)-based attacks.
But in order to succeed, SAT-based attacks implicitly assume a \emph{perfectly} reverse-engineered circuit, which is difficult to achieve in practice due to reverse engineering (RE) errors caused by automated circuit extraction.
In this paper, we analyze the effects of random circuit RE-errors on the success of SAT-based attacks.
Empirical evaluation on ISCAS, MCNC benchmarks as well as a fully-fledged \texttt{RISC-V CPU} reveals that the attack success degrades exponentially with increase in the number of random RE-errors.
Therefore, the adversaries either have to equip RE-tools with near perfection or propose better SAT-based attacks that can work with RE-imperfections.

\end{abstract}


\input{introduction.tex}

\input{sat-attack.tex}

\input{motivation.tex}

\input{proposed.tex}

\input{results.tex}
\input{conclusion.tex}


{\fontsize{16}{16}\selectfont{\bibliographystyle{IEEEtran}}
\bibliography{refs}}

\end{document}

%% file: introduction.tex
\vspace{-.5em}
\section{Introduction}
\vspace{-0.5em}
\label{sec:introduction}
\noindent

Logic locking has been proposed to mitigate hardware IP-theft, which is a serious concern for the IC industry~\cite{farinaz:epic,jv:dac12,dupuis:iolts14,jv:tc15,SARLock,Xie:2019,cyc-obfuscation,TTLock,SFLL,fortis,guin:2017,guin:2018,shift-and-leak-iccad19,rajit:encryptFF,shamsi:tifs2019,cycsat-resolvable,potluri:seql}.
Logic locking is, unfortunately, known to be vulnerable to Satisfiability (SAT)-based attacks~\cite{pramod:host15,AppSAT,meade:iscas2017,juretus:iscas2018,karmakar:iscas2018,karmakar:iscas2019,smt-attack,double-dip,lilas:aspdac19,besat,cycsat}. 
But all SAT-based attacks assume a \emph{perfectly} reverse-engineered circuit. 
This assumption may be invalid if the adversary purchases the IC from the open market, captures images and uses machine learning for automated circuit extraction, which is error prone~\cite{re-errors,re-ml,Quadir:2016}. 
There is, however, no prior work on evaluating the impact of reverse-engineering (RE) errors on attack accuracy. 
In this paper, for the first time, we analyze and quantify the effect of such errors.
The primary contributions of the paper are as follows:
\begin{itemize}

    \item We propose an error model for RE and evaluate the attack accuracy for single as well as multiple random error scenarios.
    
    \item We identify different error scenarios which can and cannot be canceled out by applying the suitable key.
    
    \item Using benchmarks as well as a \texttt{RISC-V CPU}, we quantify the impact of RE-errors on the attack accuracy. 
    
    \item Surprisingly, we unveil that in the presence of errors, increasing logic locking amount can reduce the security.
     
\end{itemize}

Our research reveals that SAT-attacks can fail even if there is a single RE-error, but there may be cases of attack success with a reasonably high error count.
The conclusions of our paper is threefold.
First, the RE tools typically need high accuracy for attacking logic locking.
Second, end-user adversary needs better SAT attacks to work with RE-errors.
Third, in majority of the cases adding more locks to the circuit increases attack success (reduces the security-level) in the presence of errors.

%% file: sat-attack.tex
\section{Satisfiability-based attack}
\label{sec:sat-attack}
\vspace{-0.5em}

SAT-solvers find the inputs to a Boolean expression such that the expression evaluates to true.
The adversary can thus formulate the locked circuit obtained through RE as a Boolean expression and use Boolean SAT-solvers to find a secret key for the given input-output pairs. 
The fundamental challenge is, however, to extract the secret key by observing a limited set of input-output pairs.
Therefore, straightforward SAT-solver usage is ineffective---although it returns a key, the key may only satisfy the limited set and not guaranteed to satisfy input-output pairs outside this set.
SAT-attack solves this problem by two intuitions.

First, unlike cryptographic constructions, the secret key itself may not be unique, causing multiple keys to form equivalence-classes that would generate identical input-output pairs.
Second, a wrong-key equivalence-class can be pruned off by determining the distinguishing input pattern (DIP), a certain input pattern which produces different outputs for two keys from two different equivalence classes.
As a result, the attack aims to perform a brute-force search on key equivalence-classes, which is more efficient than performing brute-force search on individual keys.

The attack at a high-level works as follows.
The adversary reconstructs the locked netlist through RE and formulates the circuit as a SAT-instance to find the DIPs.
The adversary then gets an activated chip from the market that can generate valid input-output pairs.
This chip is referred to as the oracle.
The adversary applies the DIP to the oracle and records the output.
The DIP-output pair is used to prune-off wrong equivalence class(es) of keys by updating the SAT-instance.
The process continues until there is only one equivalence class left, which has to be the equivalence class of the correct keys. 

\begin{algorithm}
\SetAlgoLined
\KwIn{$C$ and $eval$}
\KwResult{$\vec{K_c}$}
 $i:=1$\;
 $F_1 = C(\vec{X},\vec{K_1},\vec{Y_1}) \wedge C(\vec{X},\vec{K_2},\vec{Y_2})$\;
 \While{$sat[F_i \wedge (\vec{Y_1} \ne \vec{Y_2})]$}{
  $\vec{X^d_i} = sat\_assignment_{\vec{X}}[F_i \wedge (\vec{Y_1} \ne \vec{Y_2})]$\;
  $\vec{Y^d_i} = eval(\vec{X^d_i})$\;
  $F_{i+1} = F_i \wedge C(\vec{X^d_i},\vec{K_1},\vec{Y^d_i}) \wedge C(\vec{X^d_i},\vec{K_2},\vec{Y^d_i})$\;
  $i:=i+1$\;
 }
 $\vec{K_c} = sat\_assignment_{\vec{K_1}}(F_i)$\;
 \caption{SAT-based Logic Decryption Algorithm~\cite{pramod:host15}}
 \label{sat-attack-algo}
\end{algorithm}

%

Now we formally describe the attack.
Let the reverse-engineered locked circuit be $C(\vec{X}, \vec{K}, \vec{Y})$, which has input vector $\vec{X}$ and output vector $\vec{Y}$, and which is locked with key vector $\vec{K}$. Let the locked circuit has $M$ input bits and $L$ gates. 
Algorithm~\ref{sat-attack-algo} shows the SAT-attack in detail, where $i$ signifies the iteration number. $F_1$ is the initial SAT-instance and $F_i$ is the SAT-instance in the $i^{th}$ iteration. Each step of the algorithm is defined as follows:


\noindent {\bf Step $2$}:  Formulates the SAT-instance as two copies of the locked circuit $C(\vec{X},\vec{K_1},\vec{Y_1})$ and $C(\vec{X},\vec{K_2},\vec{Y_2})$ with same input $\vec{X}$ but different keys $\vec{K_1}$, $\vec{K_2}$ and different outputs $\vec{Y_1}$, $\vec{Y_2}$. In the next step, this formulation is exploited to generate different keys that produce different outputs. 

\noindent {\bf Step $3$}: Checks if the there are at least two different equivalence classes that satisfy the current SAT-instance. If this condition is satisfied, it enters the loop, otherwise terminates the loop; 

\noindent {\bf Step $4$}:  Runs the SAT-solver on the current SAT-instance. From the returned SAT assignment $\{\vec{X},\vec{K_1},\vec{Y_1},\vec{K_2},\vec{Y_2}\}$, this step also extracts the DIP for $i^{th}$ iteration, denoted as $\vec{X^d_i}$, 

\noindent {\bf Step $5$}: Evaluates the oracle output for $\vec{X^d_i}$, denoted as $\vec{Y^d_i}$. 

\noindent {\bf Step $6$}: Adds corresponding DIP-output constraints to the SAT-instance ($C(\vec{X^d_i},\vec{K_1},\vec{Y^d_i}) \wedge C(\vec{X^d_i},\vec{K_2},\vec{Y^d_i})$) to eliminate wrong-key equivalence-class(es) for the $i^{th}$ iteration. 

\noindent {\bf Step $9$}: Executes the SAT-solver on the SAT-instance after loop termination and extracts $\vec{K_1}$ from the returned SAT-assignment $\{\vec{X},\vec{K_1},\vec{Y_1},\vec{K_2},\vec{Y_2}\}$. Since the while condition in line $3$ fails if and only if there is a single equivalence class left, $\vec{K_1}$ is guaranteed to be {\em functionally correct}, i.e., $\vec{K_c}=\vec{K_1}$. This is a unique property of SAT-attack, which preempts verification. Otherwise even if the attacker finds the correct key, the verification step needs checking outputs for all possible $2^M$ input patterns, which is computationally infeasible for large circuits.

%% file: motivation.tex
\section{Analyzing the Impact of RE-Errors}
\label{sec:motivation}
The errors in the circuit RE can cause decryption to return the wrong keys, that is otherwise guaranteed to return the correct ones.
This section analyzes the effects of the RE errors by describing the SAT-instance formulation by using a toy example.

\begin{figure}[t]
    \centering
    \includegraphics[scale=0.38]{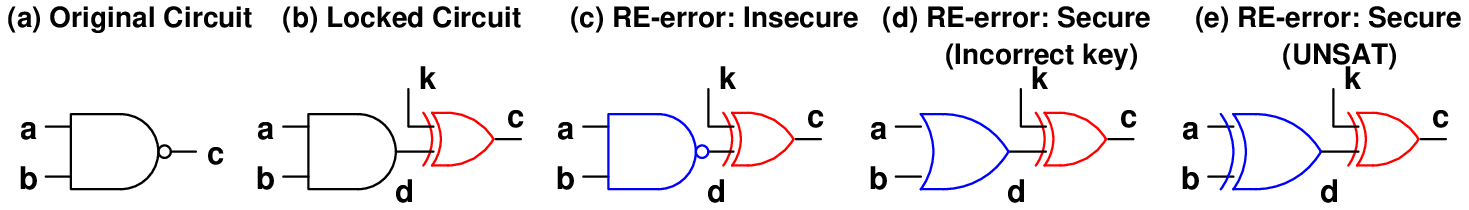}
    \caption{Motivational example to analyze impact of RE-errors}
    \vspace{-1.5em}
    \label{fig:example1}
\end{figure}

 SAT-solvers require inputs organized in a special format, called the conjunctive normal form (CNF), such that the expression evaluates to $true$.
Table~\ref{tab:sat-formulation} shows an example for the CNF formulation of $or$, $and$, $nand$, and $xor$ gates.
 Figure~\ref{fig:example1} (a) shows  a 2-bit input 1-bit output sample circuit. 
 This Boolean function takes $a, b$ as inputs and computes $c = \overline{a.b}$.  Figure~\ref{fig:example1} (b) shows the corresponding locked version with random insertion~\cite{farinaz:epic} of 1 $xor$-type key-gate (shown in red). 
 Note that the {\em correct key} in this case is `1'.
 Equations~\ref{eq:1} and~\ref{eq:2} show the conversion of the locked circuit to SAT-instances $C(\vec{X},\vec{K_1},\vec{Y_1})$ and $C(\vec{X},\vec{K_2},\vec{Y_2})$ respectively using Table~\ref{tab:sat-formulation}, where $C(.)$, the portion in black corresponds to the $and$ gate and the portion in red corresponds to the $xor$ key-gate. 
 
  \begin{multline} \label{eq:1} C(\vec{X},\vec{K_1},\vec{Y_1}) {=}  \bigg[(a+\overline{d_1})(b+\overline{d_1})(\overline{a}+\overline{b}+d_1) \color{red}(\overline{d_1}+\overline{k_1}+\overline{c_1}) \\ (d_1+k_1+\overline{c_1})(\overline{d_1}+k_1+c_1)(d_1+\overline{k_1}+c_1) \bigg]  \end{multline}  
 \begin{multline} \label{eq:2} C(\vec{X},\vec{K_2},\vec{Y_2}) {=} \bigg[(a+\overline{d_2})(b+\overline{d_2})(\overline{a}+\overline{b}+d_2) \color{red}(\overline{d_2}+\overline{k_2}+\overline{c_2}) \\ (d_2+k_2+\overline{c_2})(\overline{d_2}+k_2+c_2)(d_2+\overline{k_2}+c_2) \bigg]  \end{multline} 
Equations~\ref{eq:1} and~\ref{eq:2} are used to generate $F_1$ in step $2$ of the algorithm. Subsequently, in the first iteration, step $4$ of the algorithm returns $\{a,b,k_1,k_2,c_1,c_2\} {=} \{000101\}$, thus the first DIP is $\vec{X_1^d} {=} \{a,b\} {=} 00$ and the corresponding oracle response is $\vec{Y_1^d} {=} \overline{0.0} {=} 1$. Substituting $\vec{X_1^d}$ and $\vec{Y_1^d}$ in step $6$ yields:

\begin{equation} \label{eq:3} F_2\ {=}\ F_1  \wedge (d_1 \oplus k_1) \wedge (d_2 \oplus k_2)\ \  \end{equation} . 

Using equation~\ref{eq:3} in second iteration, the while loop in step $3$ fails, because $F_2 \wedge (c_1 \ne c_2)$ is unsatisfiable. Thus, the loop terminates and running the SAT-solver on $F_2$ (step $9$) gives $\vec{K_c} {=} 1$, which makes the locked circuit's Boolean function $(a.b) \oplus \vec{K_c} {=} (a.b) \oplus 1 {=} \overline{a.b}$. Thus, $\vec{K_c}$ is functionally correct, in the absence of RE-errors. The SAT-attack is able to find the correct key with just one DIP for this toy circuit.

\begin{table}[!t]
\begin{center}
\vspace{-0.7em}
\caption{Formulating logic gates as CNF for SAT-attack}
\vspace{-0.7em}
\begin{adjustbox}{width=\columnwidth,left}
\begin{tabular}{|c|c|}
\hline
Logic gate & Boolean clauses \\
\hline 
$z = or(x_1, x_2)$ & $ \Big(\overline{x_1} + z\Big).\Big(\overline{x_2} + z\Big).\Big(x_1 + x_2 + \overline{z}\Big)$ \\
 \hline

$z = and(x_1, x_2)$ & $\Big( x_1 + \overline{z}\Big).\Big(x_2 + \overline{z}\Big).\Big( \overline{x_1} + \overline{x_2} + z\Big)$ \\
\hline

$z = nand(x_1, x_2)$ & $\Big( x_1 + z\Big).\Big(x_2 + z\Big).\Big( \overline{x_1} + \overline{x_2} + \overline{z}\Big)$ \\
\hline

$z = xor(x_1, x_2)$ & $\Big( \overline{x_1} + \overline{x_2} + \overline{z}\Big).\Big(x_1 + x_2 + \overline{z}\Big).\Big( \overline{x_1} + x_2 + z\Big).\Big( x_1 + \overline{x_2} + z\Big)$ \\
\hline
\end{tabular}
\end{adjustbox}
\label{tab:sat-formulation}
\vspace{0.5em}
    \vspace{-2em}
\end{center}
\end{table}

\subsection{RE-Error Model}
\label{sec:error-model}
In this paper, we assume only RE-errors for 2-input logic gates. Since the possibilities are $xor,\ xnor,\ nand,\ nor,\ and$ and $or$, we consider the possible RE-errors as the possible interpretation errors between the candidate choices within this list. For example, a 2-input $nand$ gate within original netlist could be erroneously reverse-engineered as one of the candidates within the list $\{nor,\ xor,\ xnor,\ and,\ or\}$. If, e.g., it was erroneously reverse-engineered as $and$, we refer to it as $nand \rightarrow and$ type RE-error. To cover all possible cases, we also assume that the errors can occur randomly. 

Even with RE-errors, the attack can return a {\em functionally correct key}. We term them as $``$RE-errors not improving the security$"$. By contrast, for certain RE-errors (1) the attack can return a {\em functionally incorrect key}; or (2) the SAT-instance could be unsatisfiable (the SAT-solver returns {\em UNSAT}).  
We term them as $``$RE-errors improving the security$"$. Next, we discuss each of these categories in more detail.
 
\subsection{RE-errors not improving the security}
\label{sec:insecure-locking}
\noindent \textbf{Case 1:} Figure~\ref{fig:example1} (c) shows a locked circuit with $and \rightarrow nand$ type of RE-error 
with the faulty gate shown in blue. For this case, all the steps explained in the absence of error remains the same, except that in the last step, the algorithm returns $\vec{K_c} {=} 0$,  which makes the locked circuit Boolean function $\overline{(a.b)} \oplus \vec{K_c} {=} \overline{(a.b)} \oplus 0 {=} \overline{(a.b)}$. Thus, $\vec{K_c}$ is functionally correct and the \textbf{RE-error is absorbed by the key}, thus making the attack successful despite the error. This is an example of RE-error that does not improve the security. 
 
\subsection{RE-errors  improving the security}
\label{sec:secure-locking}
\noindent \textbf{Case 2:} Figure~\ref{fig:example1} (d) shows a locked circuit with $and \rightarrow or$ type of RE-error, with the faulty gate shown in blue. In this case,  

\begin{multline} \label{eq:4}  F_1 {=} \bigg[(\overline{a}+d_1)(\overline{b}+d_1)(a+b+\overline{d_1})
.\color{red}(\overline{d_1}+\overline{k_1}+\overline{c_1}) \\(d_1+k_1+\overline{c_1})(\overline{d_1}+k_1+c_1)(d_1+\overline{k_1}+c_1) \bigg] \\
\wedge \bigg[(\overline{a}+d_2)(\overline{b}+d_2)(a+b+\overline{d_2})\\
.\color{red}(\overline{d_2}+\overline{k_2}+\overline{c_2})(d_2+k_2+\overline{c_2})(\overline{d_2}+k_2+c_2)(d_2+\overline{k_2}+c_2) \bigg]\ \  \end{multline}  \\
 
\noindent Using equation~\ref{eq:4} in the first iteration, step 4 of the algorithm returns $\{a,b,k_1,k_2,c_1,c_2\}{=}000101$, thus the first DIP is $\vec{X_1^d}{=} \{a,b\} {=} 00$ and corresponding oracle output is $\vec{Y_1^d} {=} \overline{0.0}{=}1$. Substituting $\vec{X_1^d}$ and $\vec{Y_1^d}$ in step $6$ yields:

\begin{equation} \label{eq:5} F_2\ {=}\ F_1  \wedge (\overline{d_1}. k_1) \wedge (\overline{d_2}. k_2)\ \  \end{equation} . 

Using equation~\ref{eq:5} in the second iteration, the while loop in step $3$ of the algorithm fails, because $F_2 \wedge (c_1 \ne c_2)$ is unsatisfiable. Thus, the loop terminates and running SAT-solver on $F_2$ (step $9$) gives $\vec{K_c} {=} 1$, which makes the locked circuit's Boolean function $(a+b) \oplus \vec{K_c} {=} (a+b) \oplus 1 {=} \overline{a+b} {\ne} \overline{a.b}$. Thus, $\vec{K_c}$ is {\em functionally incorrect}. This is an example of RE-error that improves the security. \\

\noindent \textbf{Case 3:} Figure~\ref{fig:example1} (e) shows $and \rightarrow xor$ type of RE-error, with the faulty gate shown in blue. In this case, the SAT-attack returns {\em UNSAT}. The while loop in logic decryption algorithm terminates after first iteration because there is no key that satisfies both the original circuit as well as the DIP. This is another example of an RE-error that improves the security, because the attacker is unable to decipher the {\em correct key}.

\subsection{The Causes of the Different Cases}
\label{sec:insight}

\noindent The causes for different cases ({\em correct-key, incorrect-key, UNSAT}) for the attack with RE-errors is three-fold:
\begin{itemize}
\item The DIPs ($\vec{X^d_i}$) are generated for the locked circuit, which is erroneous; 
\item The Outputs ($\vec{Y_i}$) are evaluated on the oracle, which provides correct outputs; and 
\item The locked (erroneous) circuit is constrained to satisfy the $\{\vec{X^d_i},\vec{Y^d_i}\}$ pairs.
\end{itemize}

Because of these contradictory constraints, unlike the no-error case, in the erroneous case the algorithm is \textbf{not guaranteed} to return {\em functionally correct key}. As a result, in some cases it returns {\em UNSAT} and in some cases {\em functionally incorrect key}. This motivates us to understand the likelihood of the attack success, 
in terms of the number of cases in which the attack fails than otherwise, 
in single as well as multiple error scenarios. We therefore perform extensive experiments to evaluate the attack accuracy for wide range of error scenarios.

%% file: proposed.tex
\section{Evaluating the Impact of RE-Errors}
\label{sec:attack-accuracy-evaluation}


\begin{figure}[!t]
    \centering
    \hspace*{-3cm}   
    \includegraphics[scale=0.5,right]{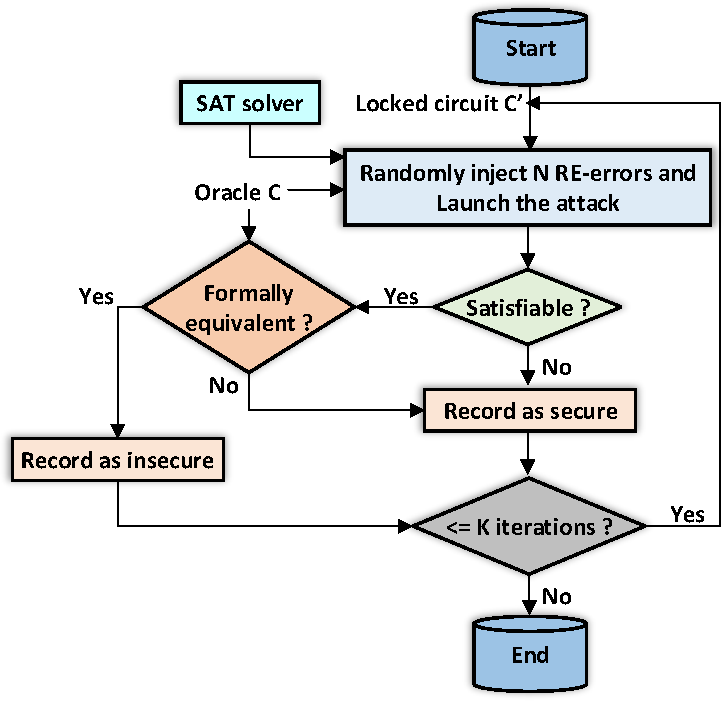}
    \vspace{-0.5em}
    \caption{Flow-chart for attack simulation with multiple-errors}
    \vspace{-1.5em}
    \label{fig:flowchart}
\end{figure}

Figure~\ref{fig:flowchart} shows the flow used to simulate SAT-attack. The only difference between conventional flow and this flow is the replacement of perfectly reverse-engineered circuit with erroneous reverse-engineered circuit. If the circuit has $L$ gates, then by definition we are allowed to inject at most $L$ errors. We inject $N \leq L$ {\em random RE-errors (multiple-error scenario)}, run the SAT-attack and subsequently perform formal-equivalence-checking if the result is satisfiable). 
The SAT-attack is recorded as {\em failed} for the cases when the result is {\em UNSAT} or when satisfiable but the decrypted key makes the locked circuit formally different from the original circuit. Otherwise it is recorded as {\em successful}. We run this procedure for $K$ random multiple-error scenarios and record the statistics. 



\subsection{Evaluation Strategy for Single and Multiple Errors}
\label{sec:single-error-model}
To systematically understand the impact of errors on the attack accuracy, we first begin with single reverse-engineering errors. We exhaustively try all gates as candidate choices, since the total simulation time is linear in the circuit size. 

Coming to multiple-errors, it is not possible to exhaust all possibilities because given an $L$-gate circuit, total number of possible multiple-error scenarios (combinations) is ${L \choose 2} + {L \choose 3} \ldots {L \choose L}$ $= 2^L - L - 1$. Since this function grows exponentially with $L$, it is not practically feasible to exhaustively evaluate all possible multiple-error scenarios for large industrial-strength circuits. Thus, we evaluate attack accuracy on a subset of random multiple-error scenarios.

%% file: results.tex
\begin{table}[!t]
\setlength{\tabcolsep}{2pt}
\begin{center}
\caption{Evaluation of Attack-Success for Various Single Reverse-Engineering Errors ($N=1$)with 5\% logic-locking.} 
\vspace{-1.8em}
\label{tab:results-single-RE-errors}
\begin{tabular}{|c|c|c|c|c|c|c|c|}
\hline 
Benchmark       &$nand \rightarrow$  
       &$xor \rightarrow$ 
       &$xor \rightarrow$ 
       &$xor \rightarrow$ 
       &$xnor \rightarrow$
       &$xnor \rightarrow$
       &$and \rightarrow$\\ 
       
       &$nor$  
       &$nor$ 
       &$xnor$ 
       &$nand$ 
       &$nand$
       &$nor$
       &$or$\\       
\hline 
\texttt{apex2}      
&$0\%$ & $71\%$ & $100\%$ & $71\%$ & $59\%$ & $13\%$ & $0.3\%$\\

\hline  
\texttt{apex4}   
&$0\%$ & $47\%$ & $100\%$ & $47\%$ & $53\%$ & $53\%$ & $0.1\%$\\

\hline 
\texttt{i4}       
&$0\%$ & $50\%$ & $100\%$ & $50\%$ & $57\%$ & $57\%$ & $0\%$\\

\hline 
\texttt{i7}      
&$0\%$ & $56\%$ & $100\%$ & $54\%$ &  $61\%$ &  $61\%$ & $0\%$ \\

\hline 
\texttt{i8}
& $0\%$ &  $38\%$ & $100\%$ & $38\%$ & $56\%$ & $56\%$ & $0.1\%$ \\

\hline 
\texttt{i9}     	
& $0\%$ & $56\%$ & $100\%$ & $56\%$ & $36\%$ & $36\%$ & $0\%$   \\

\hline 
\texttt{seq}     
& $0\%$ & $53\%$ & $100\%$ & $53\%$ & $52\%$ & $54\%$ & $0.1\%$\\

\hline 
\texttt{k2}     
& $0\%$ & $37\%$ & $100\%$ & $37\%$ & $57\%$ & $57\%$ & $0\%$ \\

\hline 
\texttt{ex1010}  
& $0\%$ &  $56\%$ & $100\%$ & $56\%$ & $58\%$ & $58\%$ & $0\%$ \\

\hline
\texttt{dalu}    
& $0\%$ & $47\%$ & $100\%$ & $48\%$ & $62\%$ & $64\%$ & $3.3\%$ \\

\hline 
\texttt{des}     
& $0\%$ & $50\%$ & $100\%$ & $51\%$ & $56\%$ & $56\%$ & $0.1\%$ \\

\hline 
\texttt{c432}   
& $0\%$ & $9\%$ & $22\%$ & $87\%$ & $67\%$ & $67\%$ & $0\%$ \\

\hline 
\texttt{c499}    
& $0\%$ & $2\%$ & $38\%$ & $2\%$ & $29\%$ & $29\%$ & $0\%$ \\
 
\hline 
\texttt{c880}    
& $0\%$ & $63\%$ & $100\%$ & $53\%$ & $64\%$ & $64\%$ & $0\%$ \\

\hline 
\texttt{c1355}   
& $0\%$ & $46\%$ & $100\%$ & $46\%$ & $38\%$ & $38\%$ & $0\%$ \\

\hline
\texttt{c1908}   
& $0\%$ & $53\%$ & $100\%$ & $53\%$ & $68\%$ & $68\%$ & $0\%$ \\

\hline 
\texttt{c2670} 
& $2\%$ & $48\%$ & $100\%$ & $48\%$ & $54\%$ & $54\%$ & $2.1\%$ \\

\hline 
\texttt{c3540}   
& $7\%$ & $53\%$ & $100\%$ & $49\%$ & $50\%$ & $50\%$ & $0\%$ \\

\hline 
\texttt{c5315}   
& $0\%$ & $35\%$ & $100\%$ & $35\%$ & $48\%$ & $48\%$ & $0\%$ \\
    
\hline 
\texttt{c7552}   
&  $1.1\%$ & $54\%$ & $100\%$ & $58\%$ &  $53\%$ & $91\%$ & $0.3\%$ \\

\hline
\texttt{RISC-V}
& $0.2\%$ & $27\%$ & $59\%$ &  $27\%$ & $35\%$ & $35\%$ & $0.1\%$ \\

\hline
\end{tabular}
\vspace{-2.8em}
\end{center}
\end{table}

\begin{figure}[!t]
\centering
\begin{tikzpicture}
\centering
\begin{axis}[
height=6.0cm,
width=0.9\columnwidth,
enlargelimits=-0.1,
y tick label style={/pgf/number format/fixed},
x tick label style={/pgf/number format/fixed},
xmax=16.00,
xmin=2.00,
ymax=30.00,
ymin=0.00,
grid=major,
legend style={at={(0.67,0.98)},anchor=north,legend columns=2,fill=none},
xlabel={$\#$ RE-Errors},
ylabel={$\%$ Attack Success },
]
\addplot [mark=xx,line width = 1.2,color=red!70!black] table[x index = 0,y index =1]{data_file_errors_attack_success.txt};
\addplot [mark=x,line width = 1.2,color=red!70!black] table[x index = 0,y index =2]{data_file_errors_attack_success.txt};
\addplot [mark=*,line width = 1.2,color=red!70!black] table[x index = 0,y index =3]{data_file_errors_attack_success.txt};
\addplot [mark=,line width=1.0,color=green!70!black] table[x index = 0,y index =4]{data_file_errors_attack_success.txt};
\addplot [mark=x,line width=1.0,color=green!70!black] table[x index = 0,y index =5]{data_file_errors_attack_success.txt};
\addplot [mark=*,line width=1.0,color=green!70!black] table[x index = 0,y index =6]{data_file_errors_attack_success.txt};
\addplot [mark=,line width=1.0,color=blue!70!white] table[x index = 0,y index =7]{data_file_errors_attack_success.txt};
\addplot [mark=x,line width=1.0,color=blue!70!white] table[x index = 0,y index =8]{data_file_errors_attack_success.txt};
\addplot [mark=*,line width=1.0,color=blue!70!white] table[x index = 0,y index =9]{data_file_errors_attack_success.txt};
\addplot [mark=,line width=1.0,color=red!70!white] table[x index = 0,y index =10]{data_file_errors_attack_success.txt};
\addplot [mark=x line width=1.0,color=red!70!white] table[x index = 0,y index =11]{data_file_errors_attack_success.txt};
\addplot [mark=*,line width=1.0,color=red!70!white] table[x index = 0,y index =12]{data_file_errors_attack_success.txt};
\addplot [mark=,,line width=1.0,color=yellow!70!black] table[x index = 0,y index =13]{data_file_errors_attack_success.txt};
\addplot [mark=x,line width=1.0,color=yellow!70!black] table[x index = 0,y index =14]{data_file_errors_attack_success.txt};
\addplot [mark=*,line width=1.0,color=yellow!70!black] table[x index = 0,y index =15]{data_file_errors_attack_success.txt};
\addplot [mark=,line width=1.0,color=green!70!white] table[x index = 0,y index =16]{data_file_errors_attack_success.txt};
\addplot [mark=x,line width=1.0,color=green!70!white] table[x index = 0,y index =17]{data_file_errors_attack_success.txt};
\addplot [mark=*,line width=1.0,color=green!70!white] table[x index = 0,y index =18]{data_file_errors_attack_success.txt};
\addplot [mark=,line width=1.0,color=blue!30!black] table[x index = 0,y index =19]{data_file_errors_attack_success.txt};
\legend{\texttt{apex2}, \texttt{apex4}, \texttt{i4}, \texttt{i7}, \texttt{i8}, \texttt{i9}, \texttt{seq}, \texttt{k2}, \texttt{ex5}, \texttt{ex1010}, \texttt{dalu}, \texttt{des}, \texttt{c432}, \texttt{c499}, \texttt{c880}, \texttt{c1355}, \texttt{c1908}, \texttt{c3540}, \texttt{c5315}}
\end{axis}
\end{tikzpicture}
\vspace{-1.5em}
\caption{Increase in attack success with increase in RE-errors. $25\%$ logic-locking and $K = 1000$ are used for the evaluation.}
\vspace{-1.7em}
\label{fig:error-rate-drop}
\end{figure}
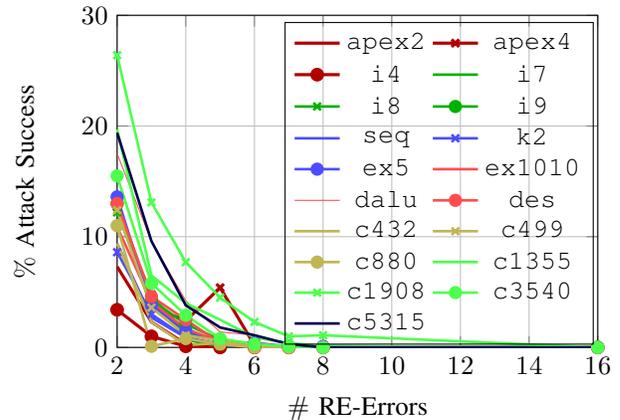

\subsection{Evaluation Results}
\label{sec:results}
 {\em IBM BladeCenter$\textsuperscript{\textregistered}$ High-Performance Cluster (HPC)} dual-core nodes with $8GB$ memory, single-threaded execution and abort limit of 1 week, are used for all the runs. 
For each value of $N$ ($\#RE-errors$), the experiments took $1$ week. We have tried for 8 values of $N$ ($2,3,4,5,6,7,8,16$), so it took altogether $8$ weeks of compute time on the {\em HPC}. 
Table~\ref{tab:results-single-RE-errors} shows the exhaustive results for single RE-errors for $8$ different types of RE-errors based on the error-model described in Section~\ref{sec:error-model}. We define \% Attack Success as $ \frac{\# insecure\  cases}{\# secure\ cases + \#\ insecure\  cases}$. 

This table shows that the attack success is a function of the type of RE-error. For e.g: for $xor \rightarrow xnor$ type RE-error, attack success is $100\%$ for most benchmarks, because the only manifestations in those cases were $xor$-type key gates, which the SAT-solver tolerates by flipping the corresponding key-bit (which is not possible for the remaining $7$ types of RE-errors). On the contrary, for $nand \rightarrow nor$ and $and \rightarrow or$ types of RE-errors, attack success was close to $0$\% for most benchmarks. For remaining types of RE-errors, the attack success was somewhere in between these two extremes. 

Figure~\ref{fig:error-rate-drop} illustrates the attack success for the different values of $N$ (X-axis) with $25$\% logic-locking and $K=1000$. This plot shows exponential degradation in attack success with linear increase in $N$. The $25\%$ logic-locking of \texttt{RISC-V CPU} using {\em sle} software~\cite{sat-attack-source-code} itself exceeds the abort limit, hence not reported. 
Figure $4$ shows the attack success as a function of \%logic locking. 
Although there is no clear trend, there is a general increase in attack success with increase in key-size. This is counter-intuitive, because we expect the attack success to degrade with increase in the key-size. 

\begin{figure}[!t]
\centering
\begin{tikzpicture}
\centering
\begin{axis}[
height=6.0cm,
width=0.9\columnwidth,
enlargelimits=-0.1,
y tick label style={/pgf/number format/fixed},
x tick label style={/pgf/number format/fixed},
xmax=15.00,
xmin=0.00,
ymax=100.00,
ymin=0.00,
grid=major,
xlabel={$\%$ Logic-Locking},
ylabel={$\%$ Attack Success},
]
\addplot [line width = 1.2] table[x index = 0,y index =1]{data_file_logic_locking.txt};
\addplot [mark=x,mark size=3.0,line width = 1.2,color=red!70!black] table[x index = 0,y index =2]{data_file_logic_locking.txt};
\addplot [mark=*,line width = 1.2,color=green!70!black] table[x index = 0,y index =3]{data_file_logic_locking.txt};
\addplot [mark=, line width=1.0,color=blue!70!white] table[x index = 0,y index =4]{data_file_logic_locking.txt};
\legend{$nand\rightarrow xor$, $xor\rightarrow nor$, $xnor\rightarrow nand$, $and\rightarrow or$}
\end{axis}
\end{tikzpicture}
\vspace{-1.0em}
\caption{Increase in \%Attack Success with \%Logic-Locking for \texttt{k2} circuit.}
\label{fig:ll-attack-success}
\vspace{-1.7em}
\end{figure}
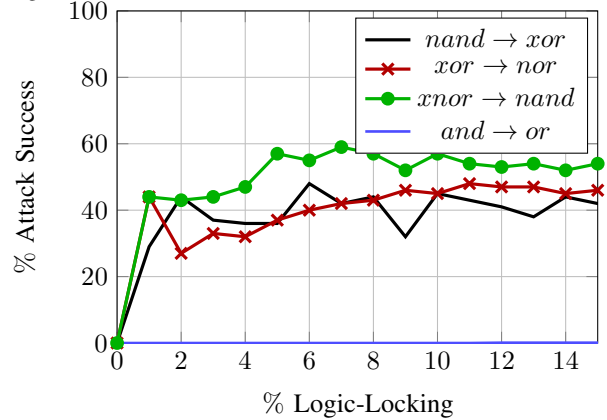

\input{removal-attack}

%% file: removal-attack.tex
\vspace{-0.3em}
\subsection{Evaluating the Error-Removal Attack}
\vspace{-0.2em}
Attack failure is registered if the SAT-instance if unsatisfiable or if satisfiable but the decrypted key verified incorrect through functional execution. In either case, the attacker's next step would be to remove RE-errors and reconstruct the original circuit. However, the attacker is unaware of the number/locations/error-scenario, so he is forced to perform brute-force checking. If there are $m$ possible error choices for each gate and $k$ RE errors, the number of possibile locations is ${L \choose k}.(m^k) $. Coming to the number of errors, $k$ can range from $1$ to $L$. So, total number of possibilities is $\sum_{k=1}^{L} {L \choose k}.(m^k)$. We know that $\sum_{k=1}^{L} {L \choose k}$ = $\mathcal{O}(2^L)$, hence the complexity of $\sum_{k=1}^{L} m^k. {L \choose k}$ is either similar or better than this. Thus error-removal and reconstruction is infeasible for large circuits. 

%% file: conclusion.tex
\vspace{-0.5em}
\section{Conclusions} 
\label{conclusion}
\vspace{-0.25em}
 In the logic-locking threat model, the adversary may not be the foundry itself but an end-user purchasing the IC from the open market.
This paper analyzes such an end-user adversary, who needs to deal with errors when doing RE of the IC. We quantify the efficacy of SAT-based attack in the presence of RE-errors and identify the underlying reasons as to why it can/cannot succeed. Empirical results suggest dependence of the attack success on the type of RE-error and  exponential degradation of attack success with error count. Surprisingly, in most cases the attack success increases with increase in key-size. 